\documentclass[twocolumn,preprintnumbers,amsmath,amsfonts,amssymb,notitlepage,showpacs,aps,prb,longbibliography,floatfix]{revtex4-2}

\usepackage{color}

\date{\today}
\usepackage{cancel}

\usepackage{amsmath}
\usepackage{graphicx}
\usepackage{hyperref}

\begin{document}
\title{The dynamical enhancement of Dzyaloshinskii-Moriya interaction in lattice Anderson impurity model
}

\author{F{\i}rat Y{\i}lmaz}
\email{firatyilmaz@enqubt.com}
\affiliation{ENQUBT, Integrated Quantum Information Technologies A.\c{S}.,  \.{I}stanbul 34467, Turkey}

\begin{abstract}
We propose a new route to generate an interband Dzyaloshinskii-Moriya interaction in metals from solely microscopic perspective. The system consists of conduction band electrons in the presence of the Rashba spin-orbit coupling which are coupled to a $f(d)$-type localized and interacting electrons. The double occupancy in the $f$-band is penalized by a significant Coulomb interaction energy, $U$. When the hybridization between two bands is not strictly local, it leads an alternative pathway to a spin Hamiltonian that may proliferate the formation of Skyrmion textures. The effective spin Hamiltonian is derived using the Schriffer-Wolff perturbation theory. In addition to a Kondo spin exchange interaction between conduction band spins and $f(d)$-band spins, which is denoted as $\vec{s}_m \cdot \vec{S}_n$, the spin Hamiltonian admits a {\bf Dzyaloshinskii-Moriya} spin coupling term, $\vec{s}_m \times \vec{S}_n$. The magnitude of the Dzyaloshinskii-Moriya term is approximately ten to hundred times smaller than that of the Kondo term under physically relevant assumptions. 

The magnitude of spin coupling terms can be engineered using time-dependent external fields, motivating a generalization to a effective time-dependent Hamiltonian. In this respect, the perturbative derivation is extended to include time-dependent external field using a time-dependent SW transformation. The form of the effective spin Hamiltonian is retained while the spin coupling coefficients gain time dependence. It has been shown that the DM term acquires a new energy scale, which leads to an enhancement of one to two orders of magnitude and makes it comparable to the Kondo coupling.
\end{abstract}

\maketitle
\section{Introduction}
Skyrmions are exotic particles which were first proposed in the context of non-linear field theories \cite{skyrme1961non}. In condensed matter systems, they are readily observed in several materials \cite{park2014observation,nagao2013direct,nagao2015real,yu2010real,jiang2017direct,wu2020observation,sohn2019real,tonomura2012real,onose2012observation}. The engineering of Skyrmions has led to a series of manipulation techniques that could be potentially used in spintronics \cite{zhang2020skyrmion}. There are indeed excellent reviews \cite{wang2022fundamental,fert2017magnetic,tokura2020magnetic} that focus on Skyrmions from broad perspectives such as theoretical and experimental characterization as well as technological applications.

Several mechanisms are known to generate Skyrmions within spin Hamiltonians, including the Dzyaloshinskii-Moriya (DM) spin interaction \cite{dzyaloshinsky1958, moriya1960anisotropic, hayami2018neel}, the double exchange mechanism \cite{azhar2017incommensurate}, and competing Heisenberg spin exchange couplings \cite{leonov2015multiply}. Such microscopic spin models often serve as effective Hamiltonians derived from more fundamental limits, like extended Hubbard or lattice Anderson models. However, derivation of these effective Hamiltonians from realistic models demands a search in a vast parameter space of candidate materials. Thereby, developing a systematic method for generating effective spin Hamiltonians from realistic models is essential. 

This work focuses primarily on the generation of Dzyaloshinskii-Moriya (DM) interactions within extended lattice Anderson models. Previous studies have extensively explored the generation of DM interactions within the same band, including topological characterization of Skyrmions \cite{hayami2018neel, bostrom2020microscopic}. Here, we investigate DM interactions between distinct electron species. Interband DM interactions have also been examined through; anisotropic effects in systems with two magnetic impurities in metals \cite{staunton1988relativistic}, SOC in effective RKKY spin Hamiltonians \cite{fert1980role}, SOC coupling between ferromagnetic layers \cite{xia1997noncollinear} and SOC of conduction electrons coupled to a single impurity \cite{zarea2012enhancement, pletyukhov2011nonequilibrium}. Notably, while the SOC in Kondo Hamiltonians is claimed not to suppress the Kondo effect (unless under an external magnetic field) \cite{meir1994spin}, it can further enhance the Kondo regime through RG analysis \cite{zarea2012enhancement}. These works listed above still lack a comprehensive approach as in the lattice Anderson impurity model, where the lattice Anderson model provides a competitive platform for the magnetic phases of materials and the Hubbard physics.

In our approach, we consider a lattice Anderson model with conduction electrons exhibiting Rashba-type SOC, coupled to flat-band $f(d)$-type electrons. This results in an effective spin Hamiltonian that includes both Kondo-type spin exchange terms, $\vec{s}_m \cdot \vec{S}_n$, and interband DM interactions, $\vec{s}_m \times \vec{S}_n$, where $\vec s_m$ and $\vec S_n$ represent the local spin moments of distinct bands on lattice sites $m$ and $n$. Our work highlights an alternative mechanism for the interband DM interactions when the hybridization potential is not strictly on-site. We use Schrieffer-Wolff perturbation theory \cite{luttinger1955motion, schrieffer1966relation} to derive the effective spin Hamiltonian while emphasizing the real-space representations, $s_m^\alpha S_n^\beta$ for physical insight.

\subsubsection{Summary of the Results}
\textit{Summary of the approach:} We first carry out a SW transformation for a time-independent Anderson lattice impurity Hamiltonian and obtain the effective spin Hamiltonian for the large Coulomb interaction, $U$.  The effective spin Hamiltonian yields two types of spin couplings, a Kondo type and a DM type. We then generalize the results to the time-dependent uniform electric field via time-dependent SW transformation. 

\textit{Time {\bf independent} results:}
The effective Hamiltonian has the following form:
\begin{eqnarray}\label{Heffective}
H_{eff} &=& \sum_{mn} J_{mn} \vec{s}_m \cdot \vec{S}_n  + \vec{D}_{mn} \cdot \left( \vec{s}_m \times \vec{S}_n \right) + \text{...}. \nonumber
\end{eqnarray}
The first term is the Kondo spin exchange term, $J_{mn}$. It is dominant at half-filling. The range of spin interactions follows the range of the hybridization potential. The second term represents the DM interaction, $\vec D_{mn}$. It is induced by Rashba-type spin orbit coupling (SOC) with magnitude $|\vec{D}_{mn}| \approx J_{mn}/10$. The remaining terms, denoted by (...), are either non-local or single-particle corrections, which we neglect as the concern is spin-spin coupling terms as well as they are suppressed at the half-filling limit.

\textit{Time {\bf dependent} results:}
The effective spin Hamiltonian retains its form under a time-dependent electric field, while the spin exchange coefficients, $J_{mn}(t), \vec{D}_{mn}(t)$ become time-dependent. The time dependence is present only for the off-site spin-spin couplings as the electric field is incorporated via Peierls phase into the system. Consequently, the relative strengths of the Kondo and the DM coupling can be tuned. It is shown that the DM term acquires a new energy scale which is one to two orders of magnitude larger than the time-independent case.

\section{The Effective Hamiltonian, $H_{eff}$}
We consider a square lattice geometry with $N$ sites. The Bravais lattice vectors are dimensionles and of unit one, denoted by $\hat e_{x,y}$. The model of concern, the periodic Anderson model includes the kinetic energy, the Coulomb interaction, the hybridization, and the Rashba-type spin-orbit coupling terms,
\begin{eqnarray}\label{Ham}
H &=& H_0 + H_1,\\
H_0 &=& H^s_{kin}+H^s_{RSC} + H^f_{kin} + H_U,\\
H^s_{kin} &=& -t_s\sum_{<ij>\sigma} c^{\dagger}_{i\sigma}  c_{j\sigma} ,\\
H^s_{RSC} &=& \lambda \sum_{mn\sigma} c^{\dagger}_{m\sigma} \left(\vec{\alpha}_{mn}\cdot \mathbb{\tau}_{\sigma \bar{\sigma}} \right) c_{n\bar{\sigma}}.,\\
H^f_{kin} &=& \epsilon^f \sum_{m\sigma} f^{\dagger}_{m\sigma} f_{m\sigma},\quad
H_U = U \sum_{m} n^f_{m\uparrow} n^f_{m\downarrow},\\
H_1 &=& t_{sf}\sum_{mn\sigma} V_{mn} c^{\dagger}_{m\sigma}f_{n\sigma} + h.c.
\end{eqnarray}  
Note that $\vec{\alpha}_{mn}^s = i t_s (\delta_m^{n+\hat{e}_y}-\delta_m^{n-\hat{e}_y},-\delta_m^{n+\hat{e}_x}+\delta_m^{n-\hat{e}_x},0)$, $n^f_{m\sigma} = f^\dagger_{m \sigma}f_{m \sigma}$ and $V_{mn}$ and $\lambda$ are dimensionless coefficients. $V_{mn}$ is the hybridization function with on- and off-site parts, $V_{mn} = V_0\delta_{mn} + V_1 \delta_{\langle mn \rangle}$. $\lambda$ is the perturbation parameter for the spin-orbit coupling.

We focus on the case where the Coulomb interaction is the largest energy scale and the system is at half-filling. The half-filling limit, $n_m^f = \sum_{\sigma}n^f_{m\sigma}= 1$ and $n^s_m = \sum_{\sigma}c^\dagger_{m \sigma}c_{m \sigma} = 1$ is achieved as the lowest energy manifold when $\epsilon^f = - U/2$ (and the chemical potential for the conduction electrons is zero). It is because $ \sum_m \epsilon^f ( n_{m\uparrow}^f + n_{m\downarrow}^f )+ U n_{m\uparrow}^f n^f_{m\downarrow} = -U/2 \sum_m (n_{m\uparrow}^f - n_{m\downarrow}^f)^2$ is minimized for half-filling. 

\subsection{Time-independent SW transformation}
A Schriffer-Wolff transformation can be applied to decouple the double occupancy subspace of $H$ from the rest.
\begin{equation}
H_{eff} = e^{S} H e^{-S} = H_0 + \frac{1}{2} [S, H_1] + \mathcal{O}(2),
\end{equation}
with the condition $[ S, H_0] =- H_1$ and $S^\dagger = -S$.

To achieve such a transformation, $S$ {\bf cannot} simply have a textbook {\bf spin-preserving form} in each fermionic species subspace, because the presence of $H^s_{RSC}$ introduces additional spin-reversing terms through $[S^{0},H^s_{RSC}]$. Therefore, we introduce two components, the {\bf spin-preserving} part, $S^{0}$ and the {\bf spin-reversing} parts, $S^1$ as $\lbrack S^{0}+ S^{1}, H_0\rbrack = - H_1$. The role of $S^{0}$ is to project $H_1$ and decouple the doubly occupant subspace, while $S^{1}$ is needed to cancel the additional spin reversing terms arising from $S^0$ via $[S^{0},H^s_{RSC}]$. The resulting the spin-preserving equation is
\begin{eqnarray}
\label{SelfCons2}
\lbrack S^{0},  H^s_{kin} + H^f_{kin} + H_U\rbrack + \lbrack S^{1},  H^s_{RSC}\rbrack   &=& -H_1,
\end{eqnarray}
as well as the spin-reversing part 
\begin{eqnarray}\label{SelfCons3}
\lbrack S^{1},  H^s_{kin} + H^f_{kin} + H_U\rbrack + \lbrack S^{0},  H^s_{RSC}\rbrack &=& 0.
\end{eqnarray}

Let us move into momentum space (for $c$ electrons) for simplicity,
\begin{eqnarray}
H^s_{kin} &=& \sum_{k\sigma} \epsilon_k c^{\dagger}_{k\sigma}  c_{k\sigma},\quad
H^s_{RSC} =\lambda \sum_{k\sigma} c^{\dagger}_{k\sigma} \left(\vec{\alpha}_{k}\cdot \mathbb{\tau}_{\sigma \bar{\sigma}} \right) c_{k\bar{\sigma}}, \nonumber\\
H_1 &=& t_{sf} \sum_{nk\sigma} V_{k} e^{-ikR_n} c^{\dagger}_{k\sigma}f_{n\sigma} + h.c.\nonumber
\end{eqnarray} 
Note that,  $\vec{\alpha}_{k}^s = -2 t_s (\sin (k_y ),-\sin (k_x),0)$, $\epsilon_k = -2t_s ( \cos (k_x) + \cos (k_y))$ and $V_k = V_0 + 2 V_1( \cos (k_x) + \cos (k_y))$ where $k_\alpha = k \cdot \hat{e}_\alpha$ as well as $k R_n = \vec k \cdot \vec{R}_n$. $R_n$ is the Bravais lattice vector for site $n$ in a 2D lattice. We neglect the arrow of $\vec k$ for notational simplicity. The $k$-space sums imply $\sum_k \equiv \frac{1}{N^{1/2}} \sum_{\vec k}$, and we consider the thermodynamic limit, $N \to \infty$ where the sums are replaced by integrals $\int \frac{d^2 k}{(2 \pi)^2}$.

The form of $S_0$ can be estimated via inspecting $[H_0,H_1]$
\begin{equation} \label{S0}
S^{0} =  t_{sf}\sum_{nk\sigma} V_k e^{-ikR_n}(A_k + B^{-}_k n^f_{j\bar{\sigma}}) c^{\dagger}_{k\sigma}f_{n\sigma} - h.c. 
\end{equation}
As discussed, $S^{1}$ becomes necessary in the presence of SOC. Inspecting $[H_{RSC}^s,H_1]$, the overall form is proposed as
\begin{equation} \label{S1}
S^{1} = -\lambda t_{sf} \sum_{nk\sigma} V_k e^{-ikR_n}(\vec{A}_k +\vec{B}_k n^f_{n\sigma})\cdot \vec{\tau}_{\sigma \bar{\sigma}} c^{\dagger}_{k\sigma}f_{n\bar{\sigma}} - h.c. 
\end{equation}
The SOC introduces a new energy scale, $\lambda t_s/U$ (in units of the Kondo coupling, $\frac{t_{sf}^2}{U}$) for $|\lambda| \ll 1$. Throughout the calculations, we keep $\lambda$ up to first order. Consequently, the second commutator in Eq.\ref{SelfCons2} is of the order $\lambda^2$ and neglected. 

Note that $A_k,B^{-}_k, \vec A_k$ and $\vec B_k$ are yet to be determined. They can be obtained when $S^{0}$ and $S^{1}$ are substituted into Eqs.\ref{SelfCons2}-\ref{SelfCons3}, which results in
\begin{eqnarray}\label{Coeffs1}
A_k &=& \frac{1}{\epsilon_k-\epsilon^f}, \quad B^{\pm}_k = \frac{1}{\epsilon_k-\epsilon^f-U}\pm\frac{1}{\epsilon_k-\epsilon^f},\\\label{Coeffs2}
\vec{ A}_k &=&  A^2_k \vec{\alpha}_{k}, \quad \vec{B}_k = B^{+}_k B^{-}_k \vec{\alpha}_{k}. 
\end{eqnarray} 
Note that the coefficients in Eq.\ref{Coeffs1} are the text-book results while the ones in Eq.\ref{Coeffs2} are due to SOC.
One can consolidate $S^0$ and $S^1$ into the following matrix form
\begin{eqnarray}
S =  t_{sf} \sum_{nk\sigma} V_k e^{-ikR_n} (A^{\sigma \sigma'}_k + B^{\sigma \sigma'}_k n^f_{n\bar{\sigma'}}&&) c^{\dagger}_{k\sigma}f_{n\sigma'} - h.c.,\nonumber\\
A^{\sigma \sigma'}_k =  \left(A_k \mathbf{1} - \lambda \vec{A}_{k} \cdot \vec{\tau}\right)_{\sigma \sigma'}, &&\quad (A^{\sigma \sigma'}_k)^\dagger =A^{\sigma \sigma'}_k,\nonumber\\
B^{\sigma \sigma'}_k = \left(B^{-}_k \mathbf{1} - \lambda \vec{B}_{k} \cdot \vec{\tau}\right)_{\sigma \sigma'},&&  \quad (B^{\sigma \sigma'}_k)^\dagger =B^{\sigma \sigma'}_k. \nonumber
\end{eqnarray} 

The form of the effective spin Hamiltonian can be obtained by evaluating the commutator $[S,H_1]$,
\begin{eqnarray}
[S,&H_1&] =  t^2_{sf} \sum_{nk\sigma \sigma'} \sum_{l k'\sigma_0} \nonumber \\
&&\lbrack V_{k} e^{-ikR_n} (A^{\sigma \sigma'}_k + B^{\sigma \sigma'}_k n^f_{n\bar{\sigma'}}) c^{\dagger}_{k\sigma}f_{n\sigma'} - h.c.,\nonumber\\
&& \quad \quad V_{k'} e^{-ikR_l}  c^{\dagger}_{k'\sigma_0}f_{l\sigma_0} + h.c. \rbrack,\\
&=& t^2_{sf} \sum_{n k k'\sigma \sigma'} V_k V_{k'}^* e^{-i(k-k')R_n}\nonumber\\ \label{Commut1}
&&\quad \quad \left( A^{\sigma \sigma'}_{k} + A^{\sigma' \sigma}_{k'} \right)c^{\dagger}_{k\sigma}c_{k'\sigma'} \\ \label{Commut2}
&&\quad + \Big[ \left( B^{\sigma \sigma'}_{k} n_{n\bar{\sigma}'}^f + B^{\sigma' \sigma}_{k'} n_{n\bar{\sigma}}^f \right)\nonumber \\
&& \quad \quad  -\left( B^{\sigma \bar \sigma'}_{k} f_{n \sigma'}^{\dagger} f_{n\bar \sigma'} +B^{\sigma' \bar \sigma}_{k'} f_{n\bar{\sigma}}^{\dagger} f_{n\sigma}  \right) \Big]c^{\dagger}_{k\sigma}c_{k'\sigma'} \\
&-&t^2_{sf} \sum_{nl k \sigma \sigma'} \mid V_k \mid^2 e^{i k R_{ln}}\label{Commut3}\\
&& \quad \left( A^{\sigma \sigma'}_{k} + A^{\sigma \sigma'*}_k 
 - \delta_{nl} \delta_{\sigma \bar \sigma'} [B^{\sigma \bar \sigma}_{k} + B^{\sigma \bar \sigma *}_k]\right) f^{\dagger}_{l\sigma}f_{n\sigma'}\nonumber \\ 
&& \quad  + \left( B^{\sigma' \sigma}_{k} n_{l\bar{\sigma}}^f + B^{\sigma' \sigma*}_{k} n_{n\bar{\sigma}'}^f  \right) f^{\dagger}_{l\sigma}f_{n\sigma'}\label{Commut4}.
\end{eqnarray}
$R_{ln}$ is $R_{l}-R_{n}$. Here, we neglect $f$-band-induced electron-electron interactions, i.e. possible pairing terms such as $ \sim - t_{sf}^2 V_k V_{k'} B_k^{\sigma \sigma'} c^\dagger_{k \sigma} c^\dagger_{k' \sigma'} f_{n \bar \sigma'} f_{n \sigma'} $.

The first sum in $[S, H_1]$ contains $c$-$c$ type couplings (Eq.\ref{Commut1}) and $cc$-$ff$ type couplings (Eq.\ref{Commut2}). The latter defines second-order processes represented by $PQQP$, where $P$ projects the $f$-electrons onto the non-doubly occupied sector and $Q = \mathbf{1} - P$ projects onto the doubly occupied sectors of the $f$-band. The second summation involves purely $f$-$f$ type couplings. The diagonal part of $A^{\sigma \sigma'}_k$ dresses the kinetic energies of the $c$-type fermions in Eq. \ref{Commut1} and introduces mobility to the $f$-electrons in Eq.\ref{Commut3}. Additionally, Eq.\ref{Commut4} lives in the single doubly occupied high-energy subspace and is dropped.

We focus on the spin-exchange terms that are given in Eq.\ref{Commut2}. It provides two types of spin-spin interactions: the Kondo ($\vec{s}_m \cdot \vec{S}_n$) and the Dzyaloshinskii-Moriya ($\vec{s}_m \times \vec{S}_n$) couplings. 

Physical visualization is difficult in the momentum space, the final expressions are thereby presented in real space. For this purpose, we use the matrix notation
\begin{align}\label{realSpace}
	\begin{pmatrix}A^{\sigma \sigma'}_{mn}\\ B^{\sigma \sigma'}_{mn} \end{pmatrix} &=&\begin{pmatrix}  A_{mn}\mathbf{1} - \lambda \vec A_{mn} \cdot \vec{\tau}_{\sigma \sigma'}\\
	 B^{-}_{mn} \mathbf{1} - \lambda \vec B_{mn} \cdot \vec{\tau}_{\sigma \sigma'} \end{pmatrix},
\end{align}
as well as the following inverse Fourier transforms,
\begin{equation}
\label{realSpace2}
\begin{pmatrix}V_{mn}\\A_{mn}\\ B^{-}_{mn}\\\vec{A}_{mn}\\ \vec{B}_{mn} \end{pmatrix} = \sum_k e^{ikR_{mn}}V_k \begin{pmatrix}1\\A_{k}\\ B^{-}_{k} \\ A^2_{k} \vec{\alpha}_k\\ B^{-}_{k}B^{+}_{k} \vec{\alpha}_k\end{pmatrix}.
\end{equation}

Considering the on-site terms for the conduction electrons, $c^\dagger_{m\sigma} c_{m\sigma'}$ and exploiting the half-filling limit, $n^f_{n\bar\sigma} = 1 - n^f_{n\sigma}$, one can then rewrite $cc$-$ff$ type couplings of $[S,H_1]$ in Eq.\ref{Commut2} as
\begin{eqnarray}
[S,H_1] &=&  -t^2_{sf} \sum_{mn \sigma \sigma'}\Bigg( V_{mn}^* \left( B^{\sigma \sigma'}_{mn} n_{n\sigma'}^f  + B^{\sigma \bar{\sigma}'}_{mn} f_{n\sigma'}^{\dagger} f_{n\bar{\sigma}'}\right) \nonumber\\
+&V_{mn}&\left( B^{\sigma' \sigma}_{mn} n_{n\sigma}^f + B^{\sigma' \bar{\sigma}}_{mn} f_{n\bar{\sigma}}^{\dagger} f_{n\sigma}  \right) \Bigg)c^{\dagger}_{m\sigma}c_{m\sigma'}\nonumber + (...) \\
= -t^2_{sf} & \sum_{mn}& V_{ mn}^*
\text{Tr}\Big[(B^-_{mn}-\lambda \tilde B_{mn} \vec \alpha_{mn}\cdot \vec \tau)  \nonumber \\ &\times& (\frac{1}{2} + \vec{S}_n\cdot \vec \tau)(\frac{1}{2} + \vec{s}_m\cdot \vec \tau)\Big]\quad+ h.c. + (...), \label{commutator}\\
=  -4t^2_{sf} & \sum_{mn}& \big( \Re{(V_{mn}^* B^-_{mn})} \vec s_m \cdot \vec S_n \nonumber\\
&& + \lambda \Re{(V_{nj}^* i \vec B_{mn})} \cdot(\vec s_m \times  \vec S_n)\big) + (...),\label{FinalTrace}
\end{eqnarray}
where we collected only the spin-spin couplings and the remaining single-particle terms are denoted by $(...)$. The complex conjugated forms are kept for the time-dependent case. In the time-independent case, the expressions within $\Re$ are readily real valued.

\subsection{Time-independent Effective Spin Hamiltonian}
One can rearrange Eq.\ref{FinalTrace} to obtain the {\bf effective Hamiltonian} as
\begin{equation}
\label{HeffectiveTimInd}
H_{eff} = \sum_{mn} J_{mn} \vec{s}_m \cdot \vec{S}_n + \vec{D}_{mn} \cdot \left( \vec{s}_m \times \vec{S}_n \right) + \text{...}
\end{equation}
where $J_{mn}$ is the Kondo interaction, $\vec D_{mn}$ is the DM interaction terms. The range of $J_{mn} = J_{m-n} $ and $\vec{D}_{mn}=\vec{D}_{m-n}$ are strictly determined by the hybridization potential. We consider on-site and nearest-neighbor hybridization terms, given by $V_{mn} = \delta_{mn} + V_1 \delta_{\langle mn\rangle}$. The remaining terms, denoted by (...), are either corrections to $H_0$ or the nonlocal terms (e.g. the constrained kinetic term), which despite having a comparable energy scale, are omitted at the strict half-filling limit. 

The first notable term is the Kondo term, $J_{mn} = J_{m-n}  = - 2 t_{sf}^2 V_{mn} B_{mn}^-$. Its range depends on the hybridization cutoff. 
Expanding the integral for $B^-_{mn}$ in Eq.\ref{realSpace2} assuming $\epsilon^f = -U/2$ and the large-$U$ limit, we obtain the following approximation
\begin{equation}
    J_{mn} = \frac{4t_{sf}^2}{U} 2V_{mn}^2.\label{KondoCoeff}
\end{equation}

The second term is the DM interaction, where $\vec{D}_{mn} = \vec{D}_{m-n} = - 2 t_{sf}^2 \lambda V_{mn}  i \vec{B}_{mn}$. Its energy scale is proportional to the Rashba spin orbit coupling parameter, $\lambda$. Expanding the integral for $\vec B_{mn}$ in Eq.\ref{realSpace2} to the first non-vanishing order in the large-$U$ limit, 
\begin{eqnarray}
    \vec D_{mn} &\approx&  \lambda \frac{4 t_{sf}^2}{U} \frac{t_s^2}{U^2} 2^{5} V_{mn} \frac{i}{t_s}\Big[ \label{DMCoeffCancel} V_0  \sum_{r \in \{\pm \hat e_x, \pm \hat e_y\}} \vec \alpha_{m,n+r} \\
    &&\quad \quad \quad  + V_1 \sum_{r,s \in \{\pm \hat e_x, \pm \hat e_y\}} \vec \alpha_{m,n+r+s}\Big], \label{eq:secondSum}\\
    &=&  \lambda \frac{4 t_{sf}^2}{U} \frac{t_s^2}{U^2} 2^{7} V_{mn}^2 \frac{i \vec{\alpha}_{mn}}{t_s}. \label{DMCoeff}
\end{eqnarray}
The first sum above in Eq.\ref{DMCoeffCancel} might look non-vanishing for $\vec D_{mm}$, but $\sum_r \vec \alpha_{m,m+r} = \sum_r \vec \alpha_{-r,0} = 0$. Aside, the DM term must have time-reversal and four-fold rotational symmetries, which could also be used to argue for a vanishing on-site DM term. Aside, another case where $D_{mn}$ vanishes is when $U\to \infty$ at the strict half filling limit due to the particle-hole symmetry. The second sum in Eq.\ref{eq:secondSum} is non-vanishing for $r=-s$. Consequently, a nonzero DM term is possible only for momentum-dependent hybridization between $c-f$ electrons, i.e. a finite $V_1$ term. There are candidate materials with momentum-dependent hybridization, which are referred to as $(115)$-compounds such as CeXIn$_5$ with X: Co, Ru, Ir \cite{burch2007optical}. The various aspects of momentum-dependent hybridization are investigated \cite{ghaemi2007higher} with a slave boson model. 

In order to obtain a rough estimate on the strength of the DM term compared to Kondo term, let us take their ratio,
\begin{equation}
    \frac{|\vec D_{mn}|}{|J_{nn}|} \approx 2^6 \lambda \left( \frac{t_s}{U}\right)^2\left( \frac{V_1}{V_0}\right)^2.\label{JOverD}
\end{equation}
Consider $U = -2 \epsilon^f \gg t_s \gg t_{sf} \sim \lambda t_s$ and $V_0 \gg V_1$. If one assumes that for each scale $ a \gg b$, we set  
\begin{eqnarray}
a = 5 b \quad &\rightarrow &\quad  J_{nn} \approx 0.02 |\vec{D}_{\langle m,n \rangle}|, \label{CaseRatio1}\\
a = 3 b \quad &\rightarrow& \quad J_{nn} \approx 0.25 |\vec{D}_{\langle m,n \rangle}|.
\end{eqnarray}

In the mean-field limit for $f$-type electrons (Hund's limit), $\vec S_i$ acts as a classical local moment, with the conduction band spins aligned according to their neighboring average local moment. Moreover, the local moments are modified as follows: $\vec{S}_i \to \vec{S}_i + \beta \vec{S}_\perp^{avg.}$ where $\vec{S}_\perp$ arises due to the spin-orbit coupling of the conduction band and proportional to Eq.\ref{JOverD}. This could induce spiral spin textures for the conduction electrons. We underline that the current model does not yield a spin-spin correlation neither in the conduction or $f$-electrons only, e.g. $\vec{S}_i \cdot \vec{S}_j$. Even if it is not the main focus of the current work, one route for such a term could be to grant mobility to the $f$-electrons or introduce a Hubbard interaction between the conduction electrons. Because a typical time-scale for the $f$-band electrons is considered as much slower than the conduction electrons meanwhile the conduction electrons are considered as a Fermi liquid, we simply neglect these two terms, which could very well be incorporated.

\section{the Effective Hamiltonian $H_{eff}(t)$: a Time dependent external field}

It is desirable to tune the spin-exchange parameters. A straightforward way to achieve such an tuning could be via external fields. Consider a uniform and time-dependent electric field, $\vec E_t = -\partial_t \vec{A}_t$ incident on the sample, which is positioned on the $x-y$ plane. The corresponding effect is imprinted as the Peierls phase, which is given by,
\begin{equation}\label{PeierlsPhase}
\phi_{mn,t} = \int_{\vec{R}_n}^{\vec{R}_m} \vec{A}_t \cdot d\vec{l} = \vec{\phi}^0_t\cdot \vec{R}_{mn}. 
\end{equation} 
Clearly, the polarization of the electromagnetic field, or the direction of the applied electric field determines the form of $\vec{\phi}^0_t = (\phi_{xt}^0,\phi_{yt}^0,\phi_{zt}^0)$.
The field is assumed to be adiabatically varying in time. Beware that time-dependent terms are denoted with an additional subscript '$t$' to distinguish them from the time-independent case. 

To derive the effective Hamiltonian, one can employ a time-dependent Schriffer-Wolff transformation \cite{eckstein}. The time-dependent lattice Anderson impurity Hamiltonian is given as
\begin{eqnarray}\label{HamPeierls}
H^s_t &=& -t_s\sum_{<mn>\sigma} e^{i\phi_{mn,t}}c^{\dagger}_{m\sigma}  c_{n\sigma} + h.c. \nonumber\\ &=&\sum_{k\sigma} \epsilon_{kt} c^{\dagger}_{k\sigma}  c_{k\sigma},\\
H^s_{RSC,t} &=& \lambda \sum_{mn\sigma} e^{i\phi_{mn,t}} c^{\dagger}_{m\sigma} \left(\vec{\alpha}_{mn}\cdot \mathbb{\tau}_{\sigma \bar{\sigma}} \right) c_{n\bar{\sigma}} + h.c. \nonumber \\
&=&\lambda \sum_{k\sigma} c^{\dagger}_{k\sigma} \left(\vec{\alpha}_{kt}\cdot \mathbb{\tau}_{\sigma \bar{\sigma}} \right) c_{k\bar{\sigma}},\\
H_{1t} &=& t_{sf}\sum_{mn\sigma} V_{mn} e^{i\phi_{mn,t}} c^{\dagger}_{m\sigma}f_{n\sigma} + h.c.  \nonumber\\
&=&t_{sf} \sum_{nk\sigma} V_{kt} e^{-ikR_{n}} c^{\dagger}_{k\sigma}f_{n\sigma} + h.c. \label{Vkt}
\end{eqnarray} 
where $H_{kin}^f$ and $H_U$ parts are on-site, and thereby time-independent and the remaining terms acquire a  Peierls phase.

The modified parameters are,
\begin{eqnarray}
\epsilon_{kt} &=& -2t_s \left( \cos (k_x-\phi_{xt}^0) + \cos (k_y-\phi^0_{yt}) \right),\nonumber\\
\vec{\alpha}_{kt} &=& -2t_s (\sin (k_y - \phi_{xt}^0),-\sin (k_x -\phi_{yt}^0),0),\nonumber\\
\quad V_{kt} &=& V_0 + 2 V_1 \cos{(k_x-\phi^0_{xt})}+ 2V_1 \cos{(k_y-\phi^0_{yt})}.\nonumber
\end{eqnarray} 
Uniformity requires an equal field strength in both direction, $\phi_{xt}^0 = \phi_{yt}^0 = -\phi_{-xt}^0=\phi_{-yt}^0= \phi^0_t$.

Consider a time-dependent operator $S_t = - S_t^\dagger$. The SW transformation now has an additional term as can be seen below,
\begin{eqnarray}
    e^{S_t} H | \Psi(t) \rangle &=& e^{S_t}i \partial_t | \Psi(t) \rangle \nonumber\\    
    e^{S_t}(H + i \partial_t S_t) e^{-S_t} | \Psi'_t \rangle &=& i  \partial_t | \Psi'_t \rangle \\    
    H_{eff,t} =e^{S_t}(H + i \partial_t S_t) e^{-S_t}&,&
    | \Psi'_t \rangle = e^{S_t}| \Psi_t \rangle.
\end{eqnarray}
Expanding $H_{eff,t}$ using Baker–Campbell–Hausdorff formula,
\begin{eqnarray}
H_{eff,t} &=& H_{0t} + H_{1t} + i\partial_t S_t + [S_t,H_{0t}+H_{1t}+i\partial_t S_t] \nonumber\\
&&\quad + \frac{1}{2} [S_t,[S_t,H_{0t}]] + \mathcal{O}(3) \label{BKHHeff_t}
\end{eqnarray}
one can eliminate the $c$-$f$ couplings in the quadratic order by the matching condition $[S_t,H_{0t}] = -H_{1t}-i\partial_t S_t $. Keeping the terms up to second order, we obtain
\begin{equation} \label{HeffApprox}
    H_{eff,t} \approx H_{0t} + \frac{1}{2} [S_t,H_{1t}+i\partial_t S_t].
\end{equation}
It is clear that the time-dependent effective Hamiltonian acquires the additional term $\sim [S_t,i\partial S_t]$, which will be examined in Sec.\ref{subsec:SdtS} exclusively. The remaining section focuses solely on the derivation of the $S_t$ operator.

Inspired by the time-independent operator $S$, let us propose the following ansatz for $S_t$,
\begin{eqnarray}
S_t =  t_{sf} \sum_{nk\sigma\sigma'} e^{-ikR_n} (&A^{\sigma \sigma'}_{kt}& + B^{\sigma \sigma'}_{kt} n^f_{n\bar{\sigma'}}) c^{\dagger}_{k\sigma}f_{n\sigma'} - h.c.,\nonumber\\\label{Stt}\\\label{Akss}
A^{\sigma \sigma'}_{kt} &=& A_{kt}\mathbf{1} - \lambda \vec{A}_{kt} \cdot \vec{\tau}_{\sigma \sigma'},\\\label{Bkss}
B^{\sigma \sigma'}_{kt} &=& B^-_{kt}\mathbf{1} - \lambda \vec{B}_{kt}  \cdot \vec{\tau}_{\sigma \sigma'} .
\end{eqnarray} 
where only the undetermined coefficients are time-dependent. $S_t$ has again the spin-preserving and the spin-reversing parts. 
\begin{eqnarray}
S_{t}^0 &=&  t_{sf} \sum_{nk\sigma} e^{-ikR_n} (A_{kt} + B^-_{kt} n^f_{n\bar{\sigma}}) c^{\dagger}_{k\sigma}f_{n\sigma} - h.c. \\
S_{t}^1 &=& -\lambda t_{sf} \sum_{nk\sigma} e^{-ikR_n} (\vec A_{kt} + \vec B_{kt} n^f_{n\sigma})\cdot \vec \tau_{\sigma \bar \sigma} c^{\dagger}_{k\sigma}f_{n\bar \sigma} - h.c. \nonumber \\
\end{eqnarray}
Unlike the time-independent case as in Eqs.\ref{S0}-\ref{S1}, $V_{kt}$ is now incorporated into the undetermined parameters of $S_t$. 

The time-dependent terms modify Eqs.\ref{SelfCons2}-.\ref{SelfCons3} as
\begin{eqnarray}\label{SelfConsT1}
i\partial_t S^{0}_t + \lbrack S^{0}_t,  H^s_{kin,t} + H^f_{kin} &+& H_U\rbrack + \lbrack S^{1}_t,  H^s_{RSC,t}\rbrack   =-H_{1t}, \nonumber\\
\label{SelfConsT2}
i\partial_t S^{1}_t + \lbrack S^{1}_t,  H^s_{kin,t} + H^f_{kin} &+& H_U\rbrack + \lbrack S^{0}_t,  H^s_{RSC,t}\rbrack = 0.
\end{eqnarray}

There are two notable works \cite{eckstein,bostrom2020microscopic} that address time-dependent external fields. In this context, the expressions we obtain consist of several convolution integrals instead of a single one, and we thereby adopt a more pedagogical approach by treating spin-reversing and -preserving parts separately. 

Although we do not explicitly present the matching equations for the time-independent parameters of $S$, $[S, H_0] =- H_1$, we currently revisit them for the time dependent case. Evaluating $[S_t, H_{0t}] =- H_1 - i\partial_t S_t$, the following set of equations are obtained,
\begin{eqnarray}\label{SelfConsEqns1}
 V_{kt} &=& (\epsilon_{kt} - \epsilon^f - i\partial_t) A_{kt},\\ \label{SelfConsEqns2}
U A_{kt}&=& (\epsilon_{kt} - \epsilon^f - U -i\partial_t) B^-_{kt},\\ \label{SelfConsEqns3}
\vec \alpha_{kt} A_{kt} &=& (\epsilon_{kt} - \epsilon^f -i \partial_t) \vec{A}_{kt},\\ 
U \vec A_{kt}+\vec \alpha_k B_{kt}^- &=& (\epsilon_{kt} - \epsilon^f -U - i \partial_t) \vec{B}_{kt}.\label{SelfConsEqns4}
\end{eqnarray} 
The partial time derivatives arising from $i\partial_t S_t$ vanish in the DC limit and we clearly obtain the expressions in Eqs.\ref{Coeffs1}-\ref{Coeffs2}. 

There are two relevant propagators required for the particular solutions,
\begin{eqnarray}
{G}_{k,1}^R(t,t') = -i \theta(t-t') &&e^{-i\int_{t'}^{t}\left(\epsilon_k(\bar{t}) -\epsilon^f - i 0^+\right) d\bar{t}},\\
{G}_{k,2}^R(t,t') = -i \theta(t-t') &&e^{-i\int_{t'}^{t}\left(\epsilon_k(\bar{t}) -\epsilon^f -U - i 0^+\right) d\bar{t}},
\end{eqnarray}
and the time-dependent parameters of $S_t$ can easily be obtained by the convolution integrals,
\begin{eqnarray}\label{timedeptCoeffModi1}
A_{kt} &=& -i\int_{-\infty}^t d t' e^{-i\int_{t'}^{t}\left(\epsilon_{k\bar t} -\epsilon^f + i 0^+\right) d\bar{t}}  V_{kt'},\\\label{timedeptCoeffModi2}
\vec{A}_{kt} &=& -i\int_{-\infty}^t d t' e^{-i\int_{t'}^{t}\left(\epsilon_{k\bar t} -\epsilon^f + i 0^+\right) d\bar{t}}  A_{kt'} \vec{\alpha}_{kt'},\\
\label{timedeptCoeffModi3}
B^-_{kt} &=& -i\int_{-\infty}^t d t' e^{-i\int_{t'}^{t}\left(\epsilon_{k\bar t} -\epsilon^f-U + i 0^+\right) d\bar{t}} U A_{kt'}\\
\label{timedeptCoeffModi4}
\vec{B}_{kt} &=& -i\int_{-\infty}^t d t' e^{-i\int_{t'}^{t}\left(\epsilon_{k\bar t} -\epsilon^f -U + i 0^+\right) d\bar{t}}  \nonumber\\
\label{timedeptCoeffModi5}
&&\quad \quad \times \left( B^-_{kt'} \vec{\alpha}_{kt'} + U \vec{A}_{kt'} \right).
\end{eqnarray}
The time dependence of the Peierls phases can take various shapes \cite{eckstein}, but in this work we assume $\phi^0_t = \phi^0 t$ for practical purposes. The evaluation of each coefficient is given in Appendix \ref{appdx:timeDepCoeffsFull} and here we provide the final approximated forms,
\begin{eqnarray}
    A_{mn,t}  &\approx&  \frac{2}{U}V_{mn} e^{-i\phi_{mn}^0 t}\left(1+\frac{2\phi_{mn}}{U} \right),\label{Amnt_apprx}\\
    B_{mn,t}^- &\approx&  - \frac{4}{U}V_{mn} e^{-i \phi_{mn}^0 t}\label{Bmnt_apprx}\\
    \vec A_{mn,t} &\approx&  \frac{4t_s}{U^2} V_0 e^{-i \phi_{mn}^0 t} \frac{\vec\alpha_{mn} }{t_s} (1+\frac{2\phi_{mn}^0}{U})\nonumber\\
    &&\quad -\frac{16t_s \phi^0}{U^3} V_1 \delta^n_m \frac{\vec \alpha_{\hat e_x}+\vec \alpha_{\hat e_y}}{t_s}\label{vecAmnt_apprx}\\
    \vec B_{mn,t} &\approx& -2^4 \frac{t_s}{U^3} \left( V_0 \phi_{mn}^0 e^{-i \phi_{mn}^0 t} + 2^3 V_1 t_s \cos (\phi^0 t)  \right) \frac{\vec \alpha_{mn}}{t_s} \nonumber\\
    &&\quad +2^6 \frac{t_s}{U^3} \delta^n_m \left(V_1 \phi^0 + 2i V_0 t_s \sin (\phi^0 t)  \right)\frac{\vec \alpha_{\hat e_x}+\vec \alpha_{\hat e_y}}{2 t_s}\nonumber\\\label{vecBmnt_apprx}.
\end{eqnarray}
It is evident that all the terms acquire a Peierls phase. The most significant change occurs in $\vec B_{mn,t}$, the emergence of a new energy scale $V_0 \phi^0_{mn}$ of the DM interaction. We further examine this new energy scale when the Kondo and the DM coupling term are calculated at the end of Sec.\ref{subsec:SdtS}.

\subsection{The time-dependent effective spin Hamiltonian} \label{subsec:SdtS}
This section aims to evaluate the commutator $[S_t, H_{1t}+i\partial_t S_t]$ and illustrate how the time-dependent $i\partial_t S_t$ renormalizes spin-couplings. The effective spin Hamiltonian in Eq.\ref{HeffectiveTimInd} retains its overall form whereas the spin exchange coefficients acquire a time dependence, denoted as $J_{mn,t}, \vec{D}_{mn,t} $. Because the time dependence within $S_t$ in Eq.\ref{Stt} are only in $A^{\sigma \sigma'}_{kt}$ and $B^{\sigma \sigma'}_{kt}$, 
\begin{eqnarray}
H_1 + i\partial S_t 
&=& \nonumber\\
t_{sf} \sum_{nk\sigma\sigma'}& e^{-ikR_n}&
 (C^{\sigma \sigma'}_{kt} + F^{\sigma \sigma'}_{kt} n^f_{n\bar{\sigma'}}) c^{\dagger}_{k\sigma}f_{n\sigma'} + h.c. \\
 C^{\sigma \sigma'}_{kt}&=& V_{kt}+ i\partial_t A^{\sigma \sigma'}_{kt}, \quad F^{\sigma \sigma'}_{kt} =i\partial_t B^{\sigma \sigma'}_{kt}. 
\end{eqnarray} 
All coefficients of the matrices are dimensionless. 
We do not explicitly evaluate the partial time derivatives here, but follow a simpler approach via working with the real-space approximated coefficients, given in Eqs.\ref{Amnt_apprx}-\ref{vecBmnt_apprx} to evaluate $i\partial_t A^{\sigma \sigma'}_{mn,t}$ and $i \partial_t B^{\sigma \sigma'}_{mn,t}$. Therefore, the operation $i\partial_t$ is carried out once the real-space expressions are obtained out of the commutator $[S_t, H_1 + i\partial_t S_t]$. Until then, we keep them as $C^{\sigma \sigma'}_{kt}$ and $F^{\sigma \sigma'}_{kt}$.
The systematic derivation, starting with $i\partial_t A^{\sigma \sigma'}_{kt}$ and $i\partial_t B^{\sigma \sigma'}_{kt}$, is necessary for completeness, which is provided in the Appendix \ref{appdx:idtSt}. 

Examining Eqs.\ref{Amnt_apprx}-\ref{vecBmnt_apprx}, $i\partial_t S_t$ acquires an overall factor which is proportional to $\sim \phi^0$, the electric field strength and vanishes for the DC limit, i.e. $i\partial_t S_t\to 0$ as $\phi^0 \to 0$. Akin to the time-independent case in Eq.\ref{commutator}, $[S_t, H_{1t}+i\partial_t S_t]$ can be evaluated as
\begin{widetext}
\begin{eqnarray}
\frac{1}{2}[S_t, H_{1t}+i\partial S_t] &=& - \frac{t^2_{sf}}{2} \sum_{mn} \nonumber \\
\Bigg[\text{Tr}\big[ B_{mn,t}(&1/2& + \vec S_n \cdot \vec \tau) (C^\dagger_{mn,t} + F^\dagger_{mn,t}) (1/2 + \vec s_m \cdot \vec \tau)\big]
+\text{Tr}\big[ B^\dagger_{mn,t}(1/2 + \vec S_n \cdot \vec \tau) (C_{mn,t} + F_{mn,t}) (1/2 + \vec s_m \cdot \vec \tau)\big]\nonumber\\
+\text{Tr}\big[ A_{mn,t}(&1/2& + \vec S_n \cdot \vec \tau) F_{mn,t}^\dagger (1/2 + \vec s_m \cdot \vec \tau)\big]
+\text{Tr}\big[ A^\dagger_{mn,t}(1/2 + \vec S_n \cdot \vec \tau) F_{mn,t} (1/2 + \vec s_m \cdot \vec \tau)\big] \Bigg]\label{wholecommutator}
\end{eqnarray}
Here, we use the matrix forms of $X_{mn,t}^{\sigma \sigma'}$ as $X_{mn,t}$ and the inverse Fourier transform for each matrix is $X_{mn,t} = \sum_k e^{i k R_{mn}} X_{kt}$. The coefficients of $C_{mn}$ and $F_{mn}$ are defined as $C_{mn}= c_{mn,t} - \lambda \vec c_{mn,t}\cdot \vec{\tau}$ and $F_{mn}= f_{mn,t} - \lambda \vec f_{mn,t}\cdot \vec{\tau}$. Assuming that the matrices has a generic from $X = x - \lambda \vec{x}\cdot \vec \tau$, trace operations given in Eq.\ref{wholecommutator} are evaluated with the following identity,
\begin{equation}
Tr[(x-\lambda \vec x\cdot \vec \tau)(1/2- \vec S_n \cdot \vec \tau)(y-\lambda \vec y\cdot \vec \tau)(1/2- \vec s_m \cdot \vec \tau)] = 2 x y \vec s_m \cdot \vec S_n - 2i\lambda (x\vec y - y \vec x)\cdot (\vec s_m \times \vec S_n) + ... \nonumber
\end{equation}
The dotted part includes the non-quartic terms and is omitted.
We are interested in the renormalized Kondo and DM coefficients within Eq.\ref{wholecommutator}, and they are readily found as
\begin{eqnarray}
J_{mn,t} &=& 2 t_{sf}^2 \Re \Big[B_{mn,t}^- (c_{mn,t}+f_{mn,t})^* + A_{mn,t} f_{mn,t}^*\Big] , \label{eq:Jmnt} \\
\vec D_{mn,t} &=& -2 t_{sf}^2 \lambda \Im \Big[i\left(B_{mn,t}^-  (\vec c_{mn,t}+ \vec f_{mn,t})^* - ( c_{mn,t}+ f_{mn,t}) \vec B_{mn,t}^*\right) + i\left( A_{mn,t} \vec f_{mn,t}^* - f_{mn,t} \vec A_{mn,t}^*\right) \Big] \label{eq:Dmnt}
\end{eqnarray} 

The coefficients of $C_{mn}$ and $F_{mn}$ are calculated via the partial time-derivatives of Eqs.\ref{Amnt_apprx}-\ref{vecBmnt_apprx},
\begin{eqnarray}
    c_{mn,t} &=& V_{mn,t} + i \partial_t A_{mn,t} =  V_{mn} e^{-i \phi_{mn}^0 t} (1+ \frac{2 \phi_{mn}^0}{U}),\nonumber \quad
    f_{mn,t} =  i \partial_t B_{mn,t}^- = - \frac{4 \phi^0}{U}V_{mn} e^{-i \phi_{mn}^0 t},\nonumber \\ 
    \vec c_{mn,t} &=&  i \partial_t \vec A_{mn,t} =  \frac{4 \phi^0 t_s}{U^2} V_0 e^{-i \phi_{mn}^0 t} \frac{\vec \alpha_{mn}}{t_s},\nonumber \quad 
    \vec f_{mn,t} =  i \partial_t \vec B_{mn,t} = - 2^4 \frac{t_s \phi_{mn}^0}{U^3} \left(V_0 \phi_{mn}^0 e^{-i \phi_{mn}^0 t}- 2^3 i V_{1} t_s \sin (\phi^0 t)  \right) \frac{\vec \alpha_{mn}}{t_s},\nonumber 
\end{eqnarray}
\end{widetext}
Substituting the findings into Eqs.\ref{eq:Jmnt}-\ref{eq:Dmnt}, the time-dependent Kondo and DM coupling terms are obtained.
\begin{eqnarray}
J_{mn,t} &\approx&  \frac{4 t_{sf}^2}{U} 2 V_{mn}^2 (1-\frac{2 \phi^0_{mn}}{U}),
\label{Jmnt}\\
\vec D_{mn,t} &\approx& \lambda \frac{4 t_{sf}^2}{U} \frac{t_s^2}{U^2} 2^{7} V_{mn}^2 \cos^2 (\phi^0 t) \frac{i \vec{\alpha}_{mn}}{t_s}\nonumber\\
&- &\lambda \frac{4 t_{sf}^2}{U} \frac{t_s \phi^0}{U^2} 2^{4} V_{mn} V_0 (1-\frac{2 \phi^0}{U})\frac{i \vec{\alpha}_{mn}}{t_s}
\label{Dmnt}
\end{eqnarray} 
There is a significant change in the DM mechanism. Even if the momentum-dependent hybridization is still necessary, a finite time-dependent electric field introduces a new energy scale into the DM term, $2 t_s V_1 \to 2t_s V_1  \cos^2 (\phi^0 t) - 2^{-3}\phi^0 V_0$. 

Let us compare the leading order factors in the Kondo and the DM couplings,
\begin{eqnarray}
    \frac{|\vec D_{mn,t}|}{|J_{mn,t}|} &\approx& \lambda\Big[ 2^6 \left( \frac{t_s}{U}\right)^2\left( \frac{V_1}{V_0}\right)^2 \cos^2 (\phi^0 t)-2^3 \frac{t_s \phi^0}{U^2}\frac{V_1}{V_0}
    \Big] \nonumber\\\label{timdepRatio}
\end{eqnarray}
There is an {\bf order of magnitude} enhancement in the DM interaction if the electric field strength, $\phi^0$ is comparable to the energy bandwidth of the conduction band, $\sim  t_s$. A more generous but relevant assumption could be $U > \phi^0 \gg t_s$, then the DM coupling is granted with an enhancement somewhere between {\bf one to two orders of magnitude}. Such an enhancement could pave the way for the DM coupling as dominant as the Kondo term.

\section{Conclusion and Outlook}
In this work, we investigated the effects of the Rashba spin-orbit coupling in metals with momentum-dependent hybridization. Our aim was to derive the effective spin Hamiltonian from a microscopic model Hamiltonian, specifically the lattice Anderson impurity model. We presented an {\bf alternative} physical mechanism to generate Dzyaloshinskii-Moriya interaction between two different bands, $\vec s_m \times \vec S_n$. We have shown that the DM coupling is not negligible compared to the Kondo coupling and gains a {\bf significant enhancement} under a time-dependent uniform electric field. The enhancement is predicted to be between one to two orders of magnitude.

The groundstate phase diagram for the effective spin interactions within the same bands is readily explored\cite{hayami2018neel,bostrom2020microscopic,rana2023skyrmions}. The extensive phase diagram of the two species of spins is currently unknown. One possible starting point could be a hybrid method, where conduction c-electrons are treated quantum mechanically, while $f$-band electrons are treated classically via a hybrid Monte-Carlo method \cite{rana2023skyrmions}. 

The local moments in the current Hamiltonian are not directly correlated due to the absence of RKKY-type terms. To establish correlations among the $\vec{S}_i$ moments, one would need to either introduce such interactions or apply an external field in the double-exchange limit. Consequently, this approach enables conduction electrons to acquire chirality around these local moments. It is important to note that a simple continuum limit does not lead to a non-linear sigma model; instead, higher-order perturbative expansions are required for each subspace. In any case, a comprehensive investigation is still necessary to fully uncover the phase diagram.

\section{Acknowledgements}
F.Y. acknowledges the early stage discussions with Arno P. Kampf on Skyrmion textures on metals.
\appendix
\begin{widetext}

\section{Time-dependent Coefficients}\label{appdx:timeDepCoeffsFull}
In this appendix, we evaluate the time-dependent coefficients in real space. Unless necessary, we take the large-$U$ limit and neglected all the coefficients, except for the electric field term, $\phi_{m-n}(t) = \delta_{\langle m,n\rangle} \phi^0(t)$ and $\epsilon_{kt}$ for $B_{mn,t}^{\sigma \sigma'}$. Note that we use a regulator for the integrals $\pm i 0^+$ incorporated into $U$ as $U^{\pm} = U \pm i 0^+$.

Throughout the calculations, we utilize following integrals
\begin{eqnarray}
    I_{m}^{\eta} (t)&=& - i\int_{-\infty}^t dt' e^{-\eta i \frac{U^{\eta}}{2} (t-t')} e^{-i \phi_{m}(t')}, \quad \eta \in \pm \\
    I_{m_1,m_2}^{\eta_1 \eta_2} (t) &=& -i\int_{-\infty}^t dt' e^{-\eta_1 i \frac{U^{\eta_1}}{2} (t-t')} e^{-i \phi_{m_1}(t')}(-i)\int_{-\infty}^{t'} dt' e^{-\eta_2 i \frac{U^{\eta_2}}{2} (t-t')} e^{-i \phi_{m_2}(t')}, \\
    I_{m_1,m_2,..,m_p}^{\eta_1 \eta_2 ... \eta_p} (t) &=& -i\int_{-\infty}^t dt_1 e^{-\eta_1 i \frac{U^{\eta_1}}{2} (t-t_1)} e^{-i \phi_{m_1}(t_1)}(-i)\int_{-\infty}^{t_1} dt_2 e^{-\eta_2 i \frac{U^{\eta_2}}{2} (t_1-t_2)} e^{-i \phi_{m_2}(t_2)} \nonumber \\
    && \quad \quad \quad ... \times (-i)\int_{-\infty}^{t_{p-1}} dt_p e^{-\eta_p i \frac{U^{\eta_p}}{2} (t_{p-1}-t_p)} e^{-i \phi_{m_2}(t_p)}.
\end{eqnarray}
We can approximate them with the following assumption, $\phi_{m-n}(t) \equiv \phi^0_{mn} t = \delta_{\langle m,n\rangle} \phi^0 t$,
\begin{eqnarray}
    I_{m}^{\eta} (t)&\approx& \eta \frac{e^{-i\phi_m^0 t}}{U/2 - \eta \phi_m^0} \approx \eta \frac{2}{U} e^{-i\phi_m^0 t} \left( 1 + \eta \frac{2 \phi_{m}}{U} \right) + O(2) \\
    I_{m_1,m_2}^{\eta_1 \eta_2} (t) &\approx& \eta_1 \eta_2 \frac{e^{-i(\phi_{m_1}^0+\phi_{m_2}^0) t}}{U/2 - \eta_1 (\phi_{m_1}^0+\phi_{m_2}^0)}  \frac{1}{U/2 - \eta_2 \phi_{m_2}^0}, \nonumber \\
    &\approx& \eta_1 \eta_2 \left(\frac{2}{U}\right)^2 e^{-i(\phi_{m_1}^0+\phi_{m_2}^0) t} \left( 1 + \frac{2}{U}[\eta_1 (\phi_{m_1}^0+\phi_{m_2}^0) + \eta_2 \phi_{m_2}^0] \right) + O(2)  \\
    I_{m_1,m_2,..,m_p}^{\eta_1 \eta_2 ... \eta_p} (t) &\approx& \eta_1 \eta_2 .. \eta_p \frac{e^{-i \sum_{q=1}^p \phi_{m_q}^0 t}}{U/2 - \eta_1 \sum_{s=1}^{p} \phi_{m_{s}}^0}  \frac{1}{U/2 - \eta_2 \sum_{s=2}^{p} \phi_{m_{s}}^0} ... \times \frac{1}{U/2 - \eta_p \phi_{m_p}^0},\\
    &\approx& \eta_1 \eta_2 .. \eta_p \left(\frac{2}{U}\right)^p e^{-i \sum_{q=1}^p \phi_{m_q}^0 t} \left( 1 + \frac{2}{U} \sum_{q=1}^p \eta_q \sum_{s=q}^{p} \phi_{m_s}^0 \right) + O(2).
\end{eqnarray}
The expansion is terminated at the first order in $\phi^0/U$.

Let us start with $A^{\sigma\sigma'}$. 
\begin{eqnarray}
A^{\sigma\sigma'} &=& A_{mn,t} - \lambda \vec A_{mn,t} \cdot \vec \tau, \nonumber\\
A_{mn,t}  
&=& \sum_{k} e^{i k R_{m-n}} (-i)\int_{-\infty}^t dt' e^{-i \frac{U^-}{2} (t-t')} V_{kt'}, \quad \text{using Eq.\ref{Vkt}}\nonumber\\
&=& V_{mn} I_{m-n}^+(t) \approx \frac{2}{U}V_{mn} e^{-i\phi_{mn}^0 t} (1+\frac{2 \phi_{mn}^0}{U}).\\
\vec A_{mn,t} &=& \sum_{k} e^{i k R_{m-n}} (-i)\int_{-\infty}^t dt' e^{-i \frac{U^-}{2} (t-t')} \vec \alpha_{kt'} (-i)\int_{-\infty}^{t'} dt'' e^{-i \frac{U^-}{2} (t'-t'')} V_{kt''},\nonumber\\
&=& \sum_{m_2\in 0 \pm \hat e_{x,y}} V_{m_2} \vec{\alpha}_{m-n-m_2} I_{m-n-m_2,m_2}^{++}(t),\nonumber\\
&=& \frac{4}{U^2}V_0 \vec \alpha_{mn}e^{-i \phi_{mn}^0 t} \left( 1 + \frac{2 \phi_{mn}^0}{U} \right) -  \delta_m^n V_1  \sum_{m_2}  \vec \alpha_{m_2} I_{-m_2 m_2}^{++}(t) \nonumber,\\
&\approx& \frac{4t_s}{U^2} V_0 e^{-i \phi_{mn}^0 t} \frac{\vec\alpha_{mn} }{t_s}\left( 1 + \frac{2 \phi_{mn}^0}{U} \right) -\frac{16t_s \phi^0}{U^3} V_1 \delta^n_m \frac{\vec \alpha_{\hat e_x}+\vec \alpha_{\hat e_y}}{t_s}
\end{eqnarray}    

The next coefficient is $B^{\sigma\sigma'}$. The $k$-space sums are done on a similar fashion and we skip them unless necessary. 
\begin{eqnarray}
B^{\sigma\sigma'}_{mn,t} &=& B_{mn,t}^- - \lambda \vec B_{mn,t} \cdot \vec \tau_{\sigma\sigma'}, \nonumber\\
B_{mn,t}^- &=& V_{mn} I_{0,m-n}^{-+}(t) \approx - V_{mn} e^{-i \phi_{mn} t}\frac{4}{U},\\
\vec B_{mn,t} &=& U \sum_{m_2}^{0, \pm \hat e_{x,y}}\sum_{m_1 }^{\pm \hat e_{x,y}} V_{m_2} \vec{\alpha}_{m_1} (-i)\int_{-\infty}^t dt' e^{i \frac{U^-}{2} (t-t')}(-i)\int_{-\infty}^{t'} dt'' \sum_k e^{i R_{m-n-m_1-m_2}}\nonumber \\
&& \quad \quad \quad \times  \left( e^{- i \int_{t''}^{t'} d\bar t (\epsilon_{k\bar t} - \frac{U^-}{2})}e^{-i \phi_{m_1}(t')}+e^{- i \int_{t''}^{t'} d\bar t (\epsilon_{k\bar t} + \frac{U^+}{2})}e^{-i \phi_{m_1}(t'')}\right) I_{m_2}^+(t'')
\\&&   \text{assume} \quad \phi_{mn}(t) = \phi^0_{mn} t = \delta_{\langle mn\rangle } \phi^0 t \quad  \text{and} \quad \int_{t''}^{t'} d \bar t\epsilon_{k\bar t} \approx -t_s \bar \epsilon_k (t'-t'')\nonumber\\
&\approx&  -2^4 \frac{t_s}{U^3} \left( V_0 \phi_{mn}^0 + 2^3 V_1 t_s \cos (\phi^0 t) e^{i \phi_{mn}^0 t } \right) \frac{\vec \alpha_{mn}}{t_s} e^{-i \phi_{mn}^0 t} \label{BmntApprox}\\
&& + 2^6 \frac{t_s}{U^3} \Big[ \delta^n_m \left(V_1 \phi^0 + 2i V_0 t_s \sin (\phi^0 t)  \right) + \delta_{\langle m,n \rangle} (...) \Big] \frac{\vec \alpha_{\hat e_x}+\vec \alpha_{\hat e_y}}{2 t_s} \\
&& + \delta_{\langle \langle m,n \rangle \rangle} (...).
\end{eqnarray}
The dotted expressions are nearest and next-nearest neighbor terms with a site independent chiral vector, $\sim (\hat e_x-\hat e_y)$. In Eq.\ref{BmntApprox}, it is evident that $\vec B_{mn,t}$  acquires orders of magnitude larger energy scale (proportional to $V_0$) under a time-dependent electric field.

\section{Derivation of $i\partial_t S_t$}\label{appdx:idtSt}
$S_t$ is given as
\begin{equation}
S_t =  t_{sf} \sum_{nk\sigma\sigma'} e^{-ikR_n} (A^{\sigma \sigma'}_{kt} + B^{\sigma \sigma'}_{kt} n^f_{n\bar{\sigma'}}) c^{\dagger}_{k\sigma}f_{n\sigma'} - h.c.
\end{equation}
Because the only time dependent terms within $S_t$ in Eq.\ref{Stt} are $A^{\sigma \sigma'}_{kt}$ and $B^{\sigma \sigma'}_{kt}$, we can obtain $i \partial_t S_t$ via substituting explicit expressions in Eqs.\ref{SelfConsEqns1}-\ref{SelfConsEqns4} into via $i\partial_t A^{\sigma \sigma'}_{kt}$ and $i\partial_t B^{\sigma \sigma'}_{kt}$.
\begin{eqnarray}
i\partial S_t 
=  t_{sf} \sum_{nk\sigma\sigma'} e^{-ikR_n}&\Big[&(\epsilon_{kt} - \epsilon^f)A^{\sigma \sigma'}_{kt} 
+ (\epsilon_{kt} - \epsilon^f-U)B^{\sigma \sigma'}_{kt}  n^f_{n\bar{\sigma'}}\Big]
 c^{\dagger}_{k\sigma}f_{n\sigma'} + h.c. \label{commutingpart}\\
+t_{sf} \sum_{nk\sigma\sigma'} e^{-ikR_n} &&(-V_{kt}+ \lambda A_{kt} \vec \alpha_{kt} \cdot \vec{\tau}_{\sigma \sigma'} + (-U A^{\sigma \sigma'}_{kt} + \lambda B^-_{kt} \vec \alpha_{kt} \cdot \vec{\tau}_{\sigma \sigma'}) n^f_{n\bar{\sigma'}}) c^{\dagger}_{k\sigma}f_{n\sigma'} + h.c.,\label{dtStt}
\end{eqnarray}
The coefficients above are dimensionless, unlike $A^{\sigma\sigma'}, B^{\sigma\sigma'} \sim \frac{1}{Energy}$. It must be underlined that when the whole commutator is considered, $[S_t, H_{1t}+i\partial_t S_t]$, the terms within $i\partial_t S_t$ in Eq.\ref{dtStt} which are proportional to $V_{kt}$ exactly cancels $H_{1t}$. This is an indication that all $V_{kt}$ terms are renormalized. For this reason, we consider the whole commutator for simplicity.

The sum in Eq.\ref{commutingpart} does not commute with $S_t$ in Eq.\ref{Stt}. $S_t$ is an anti-Hermitian and $i \partial_t S_t$ is a Hermitian operator, therefore, one cannot simply assume a vanishing commutation even for the matching parts of these two operators, $\sim [\sum_k (A + B n_{\bar \sigma'})c^\dagger f_{\sigma'} -h.c., \sum_k (...)_k(A + B n_{\bar \sigma'})c^\dagger f_{\sigma'} +h.c.] \ne 0$. The commuting part is only the imaginary part of the parameters in $i\partial S_t$, denoted by $\Im (...)_k$. Meanwhile, the real part, $\Re(...)_k$ paves the way for additional Kondo and DM coupling channels along with the additional non-matching terms. In our case, $(...)_k$ is real valued, therefore one has to consider all the terms together. We can further elaborate it if we keep $i\partial S_t$ in the following form,
\begin{eqnarray}
S_t &=&  t_{sf} \sum_{nk\sigma\sigma'} e^{-ikR_n} ( C^{\sigma \sigma'}_{kt} +  F^{\sigma \sigma'}_{kt} n^f_{n\bar{\sigma'}}) c^{\dagger}_{k\sigma}f_{n\sigma'} + h.c,\\
C_{kt}&\equiv&  C^{\sigma \sigma'}_{kt} 
=  (\epsilon_{kt} - \epsilon^f)A^{\sigma \sigma'}_{kt} + \lambda A_{kt} \vec \alpha_{kt} \cdot \vec{\tau}_{\sigma \sigma'},\\
F_{kt}&\equiv& F^{\sigma \sigma'}_{kt} 
= (\epsilon_{kt} - \epsilon^f-U)B^{\sigma \sigma'}_{kt} - U A_{kt}^{\sigma\sigma'} + \lambda B^-_{kt} \vec \alpha_{kt} \cdot \vec{\tau}_{\sigma \sigma'}.
\end{eqnarray}

Similar to Eq.\ref{commutator}, carrying out  $[S_t, H_{1t}+i\partial_t S_t]$ and recollecting the terms in the real space yields

\begin{eqnarray}
\frac{1}{2}[S_t, H_{1t}+i\partial S_t] &=& - \frac{t^2_{sf}}{2} \sum_{mn} \nonumber \\
\Bigg[\text{Tr}\big[ B_{mn,t}(&1/2& + \vec S_m \cdot \vec \tau) (C^\dagger_{mn,t} + F^\dagger_{mn,t}) (1/2 + \vec s_n \cdot \vec \tau)\big]
+\text{Tr}\big[ B^\dagger_{mn,t}(1/2 + \vec S_m \cdot \vec \tau) (C_{mn,t} + F_{mn,t}) (1/2 + \vec s_n \cdot \vec \tau)\big]\nonumber\\
+\text{Tr}\big[ A_{mn,t}(&1/2& + \vec S_m \cdot \vec \tau) F_{mn,t}^\dagger (1/2 + \vec s_n \cdot \vec \tau)\big]
+\text{Tr}\big[ A^\dagger_{mn,t}(1/2 + \vec S_m \cdot \vec \tau) F_{mn,t} (1/2 + \vec s_n \cdot \vec \tau)\big] \Bigg]
\end{eqnarray}
Each term could be evaluated with the following relation, 
\begin{equation}
Tr[(x-\lambda \vec x\cdot \vec \tau)(1/2- \vec S_m \cdot \vec \tau)(y-\lambda \vec y\cdot \vec \tau)(1/2- \vec s_n \cdot \vec \tau)] = 2 x y \vec s_n \cdot \vec S_m - 2i\lambda (x\vec y - y \vec x)\cdot (\vec s_n \times \vec S_m) + ... \nonumber
\end{equation}
The dotted part includes the non-quartic terms and is omitted. 

The coefficients of $C_{mn}$ and $F_{mn}$ are defined as $C_{mn}= c_{mn,t} - \lambda \vec c_{mn,t}\cdot \vec{\tau}_{\sigma \sigma'}$ and $F_{mn}= f_{mn,t} - \lambda \vec f_{mn,t}\cdot \vec{\tau}$
\begin{eqnarray}   
    C_{mn,t}^{\sigma\sigma'} &=& \mathcal{F}^{-1}\{(\epsilon_{kt} - \epsilon^f)A_{kt}^{\sigma\sigma'} + \lambda A_{kt} \vec \alpha_{kt}\}_{mn,t},\\
    &=& \mathcal{F}^{-1}\{\epsilon_{kt}A_{kt}^{\sigma\sigma'} \}_{mn,t} - \epsilon^f A_{mn,t}^{\sigma\sigma'} +   \lambda \mathcal{F}^{-1}\{ A_{kt} \vec \alpha_{kt} \}_{mn,t}\\
    F_{mn,t}^{\sigma\sigma'} &=&  \mathcal{F}^{-1}\{(\epsilon_{kt} - \epsilon^f-U)B_{kt}^{\sigma\sigma'}-UA_{kt}^{\sigma\sigma'} +  \lambda \mathcal{F}^{-1}\{ B_{kt}^- \vec \alpha_{kt}\}\\
    &=& \mathcal{F}^{-1}\{\epsilon_{kt}B_{kt}^{\sigma\sigma'}\}  - (\epsilon^f+U)B_{mn,t}^{\sigma\sigma'} - U A_{mn,t}^{\sigma\sigma'} + \lambda \mathcal{F}^{-1}\{ B_{kt}^- \vec \alpha_{kt} \}_{mn,t}.
\end{eqnarray}

We already obtained expressions for $A_{mn,t}^{\sigma\sigma'}$ and $B_{mn,t}^{\sigma\sigma'}$ in Appendix \ref{appdx:timeDepCoeffsFull}. The remaining task is to evaluate the inverse Fourier transforms, $\mathcal{F}^{-1}\{ A_{kt} \vec \alpha_{kt} \}_{mn,t}$ and $\mathcal{F}^{-1}\{ B_{kt}^- \vec \alpha_{kt} \}_{mn,t}$.
\begin{eqnarray}
\mathcal{F}^{-1}\{ A_{kt} \vec \alpha_{kt} \}_{mn,t}&=&\sum_{m_1m_2} V_{m_1} \vec \alpha_{m-n-m_1} e^{-i \phi^0_{m-n-m_1} t} I_{m_1}^+(t),
\\
&\approx& \frac{4t_s}{U}V_0 \frac{\vec \alpha_{mn}}{t_s} e^{-i \phi^0_{mn} t}-\frac{8 t_s}{U}V_1 \delta_m^n \left( -i \sin (\phi^0 ) + \frac{2\phi^0}{U} \cos (\phi^0 t) \right) \frac{\vec \alpha_{\hat e_x}+\vec \alpha_{\hat e_y}}{2t_s},
\\
\mathcal{F}^{-1}\{ B_{kt}^- \vec \alpha_{kt} \}_{mn,t}&=& \sum_{m_1} V_{m_1} \vec \alpha_{m-n-m_1} e^{-i \phi^0_{m-n-m_1} t} U I_{0,m_1}^{-+}(t) 
\\
&\approx& -\frac{4 t_s}{U} V_0 \frac{\vec \alpha_{mn}}{t_s} e^{-i \phi^0_{mn} t} +\frac{8 t_s}{U} V_1 \delta_m^n \frac{\vec \alpha_{\hat e_x}+\vec \alpha_{\hat e_y}}{2t_s}.
\end{eqnarray}

A similar inverse transformations could be carried out for four remaining terms, $\mathcal{F}^{-1}\{ \epsilon_{kt} A_{kt} \}_{mn,t}$, $\mathcal{F}^{-1}\{ \epsilon_{kt} \vec A_{kt} \}_{mn,t}$, $\mathcal{F}^{-1}\{ \epsilon_{kt} \vec B_{kt}^- \}_{mn,t}$ and $\mathcal{F}^{-1}\{ \epsilon_{kt} \vec B_{kt} \}_{mn,t}$. Here we provide the approximate results for $A$-terms and leave the $B$-terms to the reader.

\begin{eqnarray}
\mathcal{F}^{-1}\{ \epsilon_{kt} A_{kt} \}_{mn,t}
&\approx& -\frac{2t_s}{U}\left(\delta_{\langle mn\rangle} V_0 + 2 V_1 \delta_m^n [\cos (\phi^0 t) -i \frac{2 \phi^0}{U} \sin (\phi^0 t)]\right),\\
\mathcal{F}^{-1}\{ \epsilon_{kt} \vec A_{kt} \}_{mn,t}
&\approx& -\frac{2^4 t_s^2}{U^2} V_{mn} \frac{\vec \alpha_{mn}}{t_s} e^{-i \phi^0_{mn} t} \left( 1 + \frac{2 \phi_{mn}^0}{U} \right)  +\frac{2^4 t_s^2}{U^2} V_{0} \delta_m^n i \sin (\phi^0 t) \frac{\vec \alpha_{\hat e_x}+\vec \alpha_{\hat e_y}}{2t_s}.
\end{eqnarray}
This completes our explicit derivation for the matrix coefficients in Eq.\ref{wholecommutator}.
\end{widetext}
\bibliography{FiratSerif_DM_enhancement_arxiv}%

\end{document}